\documentclass[12pt]{mn2e}
\usepackage{graphicx}
\usepackage{epsf}
\usepackage{lscape}
\usepackage{longtable}
\usepackage{rotating}
\usepackage{color}

\newcommand{\aap}{A\&A}
\newcommand{\araa}{ARA\&A}
\newcommand{\apj}{ApJ}
\newcommand{\apjl}{ApJ}
\newcommand{\apjs}{ApJS}
\newcommand{\mnras}{MNRAS}

\begin{document}
\topmargin -0.5in 

\title[The accuracy of the UV continuum as an indicator of the SFR]{The accuracy of the UV continuum as an indicator of the star formation rate in galaxies}

\author[Stephen M. Wilkins, et al.\ ]  
{
Stephen M. Wilkins$^{1}$\thanks{E-mail: stephen.wilkins@physics.ox.ac.uk}, Violeta Gonzalez-Perez$^{2}$, Cedric G. Lacey$^{2}$, Carlton M. Baugh$^{2}$ \\
$^1$\,University of Oxford, Department of Physics, Denys Wilkinson Building, Keble Road, OX1 3RH, U.K. \\
$^{2}$\,Institute for Computational Cosmology, Department of Physics, University of Durham, South Road, Durham, DH1 3LE, U.K.\\
}
\maketitle 

\begin{abstract}
The rest-frame intrinsic UV luminosity is often used as an indicator of the instantaneous star formation rate (SFR) in a galaxy. While it is in general a robust indicator of the ongoing star formation activity, the precise value of the calibration relating the UV luminosity to the SFR ($B_{\nu}$), is sensitive to various physical properties, such as the recent star formation and metal enrichment histories, along with the choice of stellar initial mass function. The distribution of these properties for the star-forming galaxy population then suggests that the adoption of a single calibration is not appropriate unless properly qualified with the uncertainties on the calibration. We investigate, with the aid of the {\sc galform} semi-analytic model of galaxy formation, the distribution of UV-SFR calibrations obtained using realistic star formation and metal enrichment histories. At $z=0$, we find that when the initial mass function is fixed (to the Kennicutt IMF), the median calibration is $B_{\rm fuv}=0.9$ where ${\rm SFR}/[{\rm M_{\odot}\,yr^{-1}}]=B_{\nu}\times 10^{-28}\times L_{\nu}/[{\rm ergs\,s^{-1}\,Hz^{-1}}]$. However, the width of the distribution $B_{\rm fuv}$ suggests that for a single object there is around a $20\%$ {\em intrinsic} uncertainty (at $z=0$, rising to $\simeq 30\%$ at $z=6$) on the star formation rate inferred from the FUV luminosity without additional constraints on the star formation history or metallicity. We also find that the median value of the calibration $B_{\rm fuv}$ is correlated with the star formation rate and redshift (at $z>3$) raising implications for the correct determination of the star formation rate from the UV.
\end{abstract} 


\section{Introduction}

Observations of the rest-frame UV continuum of galaxies are widely used as a measure of the instantaneous star formation rate (e.g. Madau et al. 1998, Kennicutt 1998, Salim et al. 2007). UV observations are particularly important at high-redshift ($z>2$) where the rest-frame UV is shifted into the observed-frame optical and near-IR making it easily accesible to ground and space based observatories (e.g. Lilly et al. 1996, Madau et al. 1996, Bouwens et al. 2007, Bouwens et al. 2009, Wilkins et al. 2010, Wilkins et al. 2011a). This is in contrast to prominent optical emission line diagnostics, such as H$\alpha$, which are shifted beyond the $K$-band at such redshifts.

However, the use of the UV as a diagnostic has several problems. The primary shortcoming is the effect of dust attenuation, as even moderate optical attenuations can result in severe attenuation in the UV (Sullivan et al. 2001). To some extent far-IR (FIR) observations, which probe UV emission reprocessed by dust, or observations of the UV continuum slope (e.g. Bouwens et al. 2009, Wilkins et al. 2011b, Wilkins et al. 2012b, Finkelstein et al. 2012, Bouwens et al. 2012) can be used to recover the intrinsic UV luminosity\footnote{Of course both these techniques have associated problems: FIR observations generally have a much brighter flux sensitivity and are limited to low-redshift or extremely bright sources. The UV continuum slope is also sensitive to the star formation and metal enrichment histories. Sole use of the UV continuum slope plus observed UV luminosity may result in incompleteness due to heavily obscured galaxies being missed.}. 

The second principal shortcoming is that the UV-SFR calibration $B_{\nu}$, where $B_{\nu}$ is defined such that (c.f. Madau et al. 1998),
\begin{equation}\label{eqn:calib} 
{\rm SFR}/[{\rm M_{\odot}\,yr^{-1}}]=B_{\nu}\times 10^{-28}\times L_{\nu}/[{\rm ergs\,s^{-1}\,Hz^{-1}}],
\end{equation} 
where $L_{\nu}$ is the intrinsic UV luminosity and SFR is the star formation rate, is not unique but instead is sensitive to the recent star formation history, metal enrichment history and the form of the stellar initial mass function (e.g. Madau et al. 1998, Wilkins et al. 2008ab).

In this study we use the {\sc galform} semi-analytical model of galaxy formation (Cole et al. 2000, Baugh et al. 2005 - hereafter B05) to investigate how variations in the star formation and metallicity histories within a realistic population of galaxies affect the calibration, $B_{\nu}$. This paper is organised as follows: in Section \ref{sec:pp} we describe the various physical processes which affect the calibration $B_{\nu}$, including the recent star formation history (\S\ref{sec:pp.dsm.sfh}), metal enrichment (\S\ref{sec:pp.Z}) and IMF (\S\ref{sec:pp.dsm.imf}). In Section \ref{sec:gp} we use the {\sc galform} galaxy formation model to determine the distribution of $B_{\nu}$ (\S\ref{sec:gp.dist}) and investigate the correlation of the calibration with star formation rate (\S\ref{sec:gp.sfr}) and redshift (\S\ref{sec:gp.z}). Finally, in Section \ref{sec:c} we present our conclusions.

\begin{figure}
\centering
\includegraphics[width=20pc]{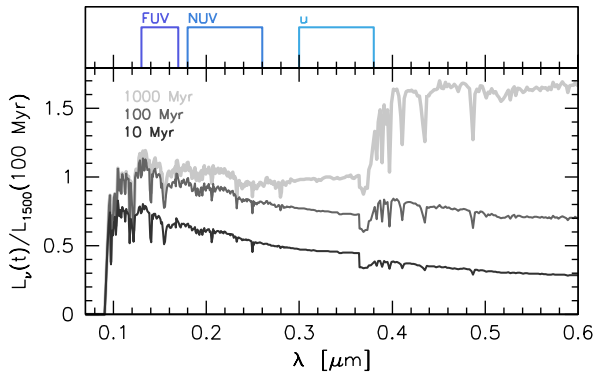}
\caption{The UV-optical SED of three composite stellar populations with constant star formation over the preceding $1000$, $100$ and $10\,{\rm Myr}$ as labelled and the three rest-frame broadband filters ($FUV$, $NUV$ and $u$) considered in this work.}
\label{fig:filters}
\end{figure}

Throughout this work we consider three artificial {\em rest}-frame UV/optical filters: a far-UV filter FUV ($T_{\lambda}=[0.13<\lambda/{\rm \mu m}<0.17]$\footnote{We utilise the Iverson bracket notation such that $[A]=1$ when $A$ is true and $0$ otherwise, i.e. all three filters have a top-hat profile.}), a near-UV one: NUV ($T_{\lambda}=[0.18<\lambda/{\rm \mu m}<0.26]$), and a {\em u}-band filter: {\em u} ($T_{\lambda}=[0.30<\lambda/{\rm \mu m}<0.38]$). The wavelength range of the FUV and NUV filters are chosen to reflect the range of the GALEX filters at $z=0$ while the {\em u}-band is chosen to cover a similar range to the SDSS $u$ and HST Wide Field Camera 3 $U_{f336w}$-bands (at $z=0$). The decision to use {\em rest}-frame filters is motivated by the desire to consistently compare the calibration at different redshifts. A top-hat profile is assumed both because it is conceptually simpler but also to allow the easy calculation of the required $k$-correction from an {\em observed}-frame filter. Fig. \ref{fig:filters} shows the three filter transmission functions together with the SEDs of three star forming galaxies (with different previous durations of star formation) for context. 

\section{Physical Properties Affecting the Calibration}\label{sec:pp}

\begin{figure}
\centering
\includegraphics[width=20pc]{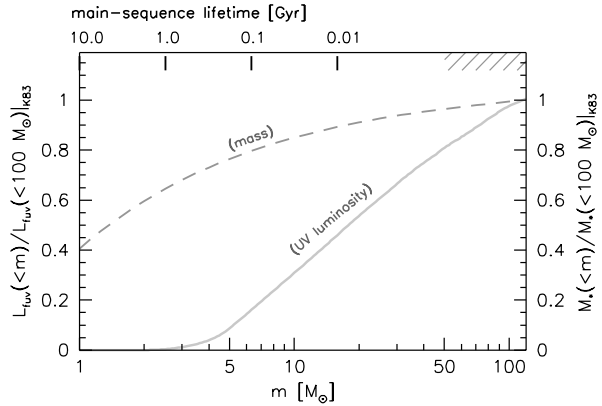}
\caption{The dashed curve shows the cumulative mass formed in stars (right-hand axis) for a Kennicutt IMF as a function of mass. The solid line shows the cumulative UV luminosity produced for this IMF (left-hand axis) as a function of limiting mass assuming a previous star formation duration of $100\,{\rm Myr}$. The main-sequence lifetimes $\tau_{\rm ms}$ (top-axis) assume the relation $\tau_{\rm ms}\approx 10^{10}\left[m/{\rm M_{\odot}}\right]^{-2.5}\,{\rm yr}$ which is approximately valid over the mass range $0.1<m/{\rm M_{\odot}}<50$.}
\label{fig:lum_mass}
\end{figure}

The stellar UV emission of a star forming galaxy is predominantly driven by high-mass stars ($m>10\,{\rm M_{\odot}}$). The short main-sequence lifetimes of these stars suggests that the UV luminosity is a potential diagnostic of the ongoing (or instantaneous) star formation rate. In reality the UV emission from a star forming galaxy is produced by stars with a range of masses, and thus main-sequence lifetimes. Fig. \ref{fig:lum_mass} shows the cumulative UV luminosity as a function of mass assuming a Kennicutt (1983) initial mass function (defined in \S\ref{sec:pp.dsm.imf}), for $100\,{\rm Myr}$ of continuous star formation, metallicity $Z=0.02$ and using the {\sc pegase.2} stellar population synthesis (SPS) code (Fioc \& Rocca-Volmerange 1997, 1999). While the most massive stars individually produce the largest UV luminosities, the sharp decline in the mass function means that the UV luminosity of a star forming stellar population is dominated by stars with $m=5-50\,{\rm M_{\odot}}$. Stars with $m<5\,{\rm M_{\odot}}$ account for $\approx 75\%$ of the mass formed but less than $10\%$ of the UV luminosity. For $m<10\,{\rm M_{\odot}}$, the fraction of mass has increased by $10\%$ but the luminosity has increased by a factor of $3$.

\subsection{Recent Star Formation History}\label{sec:pp.dsm.sfh}

\begin{figure}
\centering
\includegraphics[width=20pc]{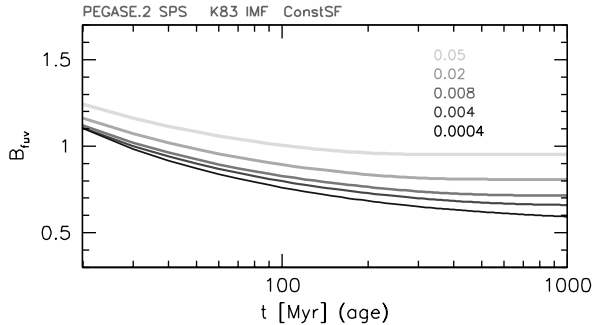}
\caption{The effect of the duration of the star formation history and the choice of metallicity on the UV-SFR calibration, $B_{\rm fuv}$. The curves show the calibration as a function of the duration of previous constant star formation (assuming a Kennicutt IMF) for 5 different metallicities, as labelled.}
\label{fig:age}
\end{figure}

The significant contribution to the UV luminosity of actively star forming galaxies of stars with $m<10\,{\rm M_{\odot}}$, which have main-sequence lifetimes $>30\,{\rm Myr}$, means that the total UV luminosity (and thus the UV-SFR calibration) is sensitive not only to the instantaneous SFR but also the {\em recent} star formation history. In Fig. \ref{fig:age} the calibration $B_{\rm fuv}$ is shown as a function of the duration of previous (constant) star formation using the {\sc pegase.2} SPS code and a Kennicutt IMF for several fixed metallicities. As the duration of preceding star formation increases, the calibration factor declines in amplitude due to the increase in $L_{\rm fuv}$ caused by the accumulation of stars with $m<10\,{\rm M_{\odot}}$ which are still luminous in the UV. After roughly $1\,{\rm Gyr}$ of continuous star formation, the additional contribution of new stars to the UV luminosity is balanced by the loss of older low-mass stars, leaving the luminosity and thus the calibration approximately constant. 

Specifically, after $100\,{\rm Myr}$ of continuous star formation with constant metallicity $Z=0.02$ and a Kennicutt (1983) IMF the {\sc pegase.2} SPS code predicts $B_{\rm fuv}=0.89$ for the $FUV$ filter (for the $NUV$ and $u$-bands we instead obtain $B_{\rm nuv}=0.97$ and $B_{\rm u}=1.16$ respectively for the same scenario). If instead the Salpeter (1955) IMF (see definition in \S\ref{sec:pp.dsm.imf}) over the mass-range $0.15-120\,{\rm M_{\odot}}$ is assumed, the {\sc pegase.2} SPS code predicts $B_{\rm fuv}=1.21$. This is similar to the Madau et al. (1998) value ($B_{1500}=1.25$) which was determined using the Bruzual \& Charlot (1993) SPS code with an updated stellar library and assuming a Salpeter IMF over the range $m=0.1-125\,{\rm M_{\odot}}$\footnote{The value $B_{\nu}=1.4$ quoted by Kennicutt (1998)  is derived in a similar way to the Madau et al. (1998) value but assumes a Salpeter IMF over $m=0.1-100\,{\rm M_{\odot}}$.}. 

\subsection{Metallicity}\label{sec:pp.Z}

The UV luminosity of a star is dependent not only on its initial mass but also on its chemical composition. Fig. \ref{fig:age} shows dependence of the calibration $B_{\rm fuv}$ on the duration of previous star formation assuming several different (fixed) stellar metallicities. At a given age, lowering the metallicity reduces the calibration $B_{\rm fuv}$ suggesting that lower metallicity stars produce a greater UV luminosity per unit stellar mass formed. This is due to the reduced effect of opacity in low metallicity stellar cores which allows high mass stars to achieve higher energy production rates and thus luminosities (and temperatures). Assuming a star formation duration of $100\,{\rm Myr}$, reducing the metallicity from $Z=0.02$ to $Z=0.004$ reduces the calibration from $B_{\rm fuv}=0.90$ to $0.79$.

\subsection{Initial Mass Function}\label{sec:pp.dsm.imf}

\begin{figure}
\centering
\includegraphics[width=20pc]{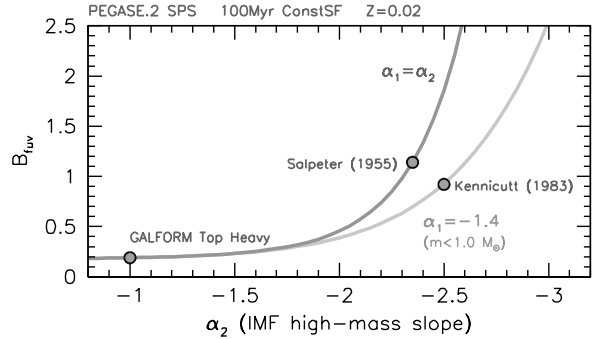}
\caption{The impact of the choice of initial mass function on the UV-SFR calibration $B_{\rm fuv}$. The two curves show the effect of changing the high-mass slope ($\alpha_2$) (upper curve: $\alpha_1=\alpha_2$, lower curve: $\alpha_1=-1.4$ with $m_{c}=1\,{\rm M_{\odot}}$). The points denote the popular Salpeter and Kennicutt IMFs. The top-heavy IMF assumed in starbursts in the default implementation of the B05 {\sc galform} model (see \S\ref{sec:gp}) is also indicated. In all cases $Z=0.02$ and there has been $100\,{\rm Myr}$ previous constant star formation.}
\label{fig:a2}
\end{figure}

The calibration is also strongly affected by the choice of stellar initial mass function (IMF). A popular representation of the IMF is a broken power-law, with the slope below some characteristic mass $m_{c}$ being flattened relative to that at high-masses. This can be written as: $\xi(m)={\rm d}N/{\rm d}m\propto m^{\alpha_1}[m_{\rm low}<m<m_c]+m^{\alpha_2}[m_c<m<m_{\rm high}]$. In Fig. \ref{fig:a2} we show how the calibration $B_{\rm fuv}$ is affected by simple changes to the IMF. The two curves show the relationship between $B_{\rm fuv}$ and the high-mass slope of the IMF for two different low mass behaviours: (1) when $\alpha_1=\alpha_2$\footnote{Which replicates the Salpeter (1955) IMF when $\alpha_1=\alpha_2=-2.35$.}, i.e. an un-broken power law, and (2) when $\alpha_1$ is fixed at $-1.4$ (with $m_c=1\,{\rm M_{\odot}}$) which replicates the Kennicutt IMF at $\alpha_2=-2.5$. In both cases $B_{\rm fuv}$ rapidly increases towards steeper high-mass slopes as the stellar mass formed becomes progressively dominated by low mass, UV faint stars. A more detailed consideration of the effect of the IMF on the recovery of physical properties such as the star formation rate, stellar mass, mass-weighted age etc. is discussed in Wilkins et al. {\em in-prep}. 

\section{Predictions from A Galaxy Formation Model}\label{sec:gp}

We use the {\sc galform} semi-analytical galaxy formation model (see Baugh 2006 for an overview of hierarchical galaxy formation models) developed initially by Cole et al. (2000) to predict the intrinsic UV properties of galaxies in a $\Lambda$CDM universe. In this paper we concentrate our attention on the Baugh et al. (2005, B05) model (see also Lacey et al. 2008, 2011) which reproduces observations of high-redshift galaxies. The B05 model uses, by default, the simple stellar population (SSP) spectral energy distributions (SEDs) generated by Bressan, Granato, \& Silva (1998), using the Padova 1994 isochrones and the model stellar atmospheres from Kurucz (1993). The defining features of the B05 model include the adoption of a top-heavy IMF ($\xi\propto m^{-1}$) in merger driven star formation (with a Kennicutt IMF for quiescent star formation) and a time-scale for quiescent star formation that is a function of the disc circular velocity, and independent of redshift. In addition to the {\em default} model we also consider two additional variants of the B05 model: (a) an implementation in which a single IMF is invoked (the Kennicutt IMF) in both quiescent and starburst modes of star formation, (b) an implementation in which both a single IMF is invoked and the stellar metallicity is fixed at $Z=0.02$. These additional implementations of the model allow us to estimate the contribution to the scatter in the calibration $B_{\nu}$ caused by the variation in the star formation and metal enrichment histories alone.

\subsection{Predicted distribution of $B_{\nu}$}\label{sec:gp.dist}

The distribution of UV-SFR calibrations, for star forming galaxies (which we take to be those with SFR$>1\,{\rm M_{\odot}\,yr^{-1}}$), predicted from the three implementations of the B05 {\sc galform} model are shown in Figs. \ref{fig:d_fuv}, \ref{fig:d_nuv}, and \ref{fig:d_u} assuming the FUV, NUV and {\em u}-band filters. The top panel of each figure shows the distribution at low-redshift ($z=0$) while the bottom panel shows the distribution at high-redshift ($z=6$). The 15.9, 50 and 84.1 percentiles of each of these distributions is also presented in Table \ref{tab:params}, along with an estimate of the fractional uncertainty. We now discuss the predictions for the three implementations of the B05 model in turn.

\begin{table}
\caption{The 15.9, 50 and 84.1 percentiles of the distribution of UV-SFR calibrations ($B_{\nu}$) and an estimate of the fractional uncertainty (defined as (P$_{15.9}$ - P$_{15.9}$)/2$\times $P$_{50}$) for both the default implementation and single IMF implementation of the B05 {\sc galform} model at $z=0$ and $z=6$. The fractional uncertainty by this definition would simply be equal to the standard deviation divided by the median were $B_{\nu}$ normally distributed. $^{a}$ in these cases the distribution is clearly non-gaussian and the fractional uncertainty is much less useful.}
\begin{tabular}{ccccccccccc}
  &  &  &  &  & fractional uncertainty \\
 $z$ & band & P$_{15.9}$ & P$_{50}$ & P$_{84.1}$ & (P$_{15.9}$ - P$_{15.9}$)/2$\times $P$_{50}$\\
\hline
\multicolumn{6}{c}{{\em default} implementation} \\
\hline
$0$ & FUV & 0.64 & 0.88 & 1.07 & 0.25 \\ 
$0$ & NUV & 0.68 & 0.91 & 1.13 & 0.25 \\ 
$0$ & {\em u} & 0.70 & 1.02 & 1.33 & 0.31 \\ 
$6$ & FUV & 0.26 & 0.30 & 0.92 & 1.10$^{a}$ \\ 
$6$ & NUV & 0.32 & 0.36 & 1.02 & 0.95$^{a}$ \\ 
$6$ & {\em u} & 0.34 & 0.42 & 1.02 & 0.81$^{a}$ \\ 
\hline
\multicolumn{6}{c}{single IMF, variable metallicity implementation} \\
\hline
$0$ & FUV & 0.73 & 0.90 & 1.09 & 0.20 \\  
$0$ & NUV & 0.74 & 0.93 & 1.17 & 0.23 \\ 
$0$ & {\em u} & 0.75 & 1.06 & 1.47 & 0.34 \\ 
$6$ & FUV & 0.79 & 1.00 & 1.39 & 0.30 \\ 
$6$ & NUV & 0.82 & 1.09 & 1.56 & 0.34 \\ 
$6$ & {\em u} & 0.82 & 1.09 & 1.52 & 0.32 \\ 
\end{tabular}
\label{tab:params}
\end{table}

\subsubsection{Single-IMF Implementation}

By considering an implementation of the B05 model in which the IMF is the same for all star formation modes we can investigate the effect of the recent star formation and metal enrichment histories on the calibration distribution. The distribution of $B_{\rm fuv}$ (shown in Fig. \ref{fig:d_fuv}) in the single IMF implementation of the model at low-redshift ($z=0$) is roughly gaussian with a median $B_{\rm fuv}=0.90$. This median value is almost exactly the same as that found simply by assuming a solar metallicity and a $100\,{\rm Myr}$ previous duration of constant star formation (as is often assumed in the literature for the determination of $B_{\rm fuv}$, e.g. Madau et al. 1998). The width of the $68.2\%$ confidence interval (i.e. the interval encompassed by the $15.9^{\rm th}$-$84.1^{\rm th}$ percentiles, P$_{84.1}-$P$_{15.9}$) is P$_{84.1}-$P$_{15.9}=0.36$ and the distribution is fairly symmetric (i.e. P$_{50}-$P$_{15.9}\approx\, $P$_{84.1}-$P$_{50}$). This implies a fractional uncertainty\footnote{This is defined as $($P$_{84.1}-$P$_{15.9})/2\times $P$_{50}$. This would simply be the standard deviation divided by the median were $B_{\nu}$ normally distributed.} on $B_{\rm fuv}$ for an individual object of around $0.2$. If instead of the FUV filter we consider the redder NUV or {\em u}-band filters (as shown in Figs. \ref{fig:d_nuv} and \ref{fig:d_u} respectively) the median calibrations increase to $B_{\rm nuv}=0.93$ and $B_{u}=1.06$ (this simply reflects that the intrinsic spectrum is blue, i.e. $\beta<-2$, as seen in Fig. \ref{fig:filters} and discussed in more detail in Wilkins et al. 2012) and the intrinsic uncertainty, at $z=0$, increases to $0.23$ and $0.34$ respectively. The increase in the scatter reflects the increasing sensitivity to the star formation history as a wider range of stellar masses contribute to the UV luminosity at these longer wavelengths.

The distribution at high-redshift ($z=6$, shown in the lower-panel of Fig. \ref{fig:d_fuv}) is similar in form though has both a larger median ($B_{\rm fuv}=1.00$) and broader confidence interval (P$_{84.1}-$P$_{15.9}=0.60$), and is skewed towards larger values of $B_{\rm fuv}$. The resulting fractional uncertainty increases to $0.30$ (with $0.34$ and $0.32$ in the NUV and {\em u}-band respectively). The consequences of the differences at high-redshift are discussed in more detail in \S\ref{sec:gp.z}.

\subsubsection{Single-IMF Single-metallicity Implementation}

To assess the relative contribution of the star formation and metal enrichment histories to the scatter in the calibration we also consider an ({\em unrealistic}) implementation of the B05 model in which the metallicity is fixed at $Z=0.02$ and a single IMF assumed. In this case the scatter in $B_{\nu}$ will be driven entirely by the variation in the star formation history. The resulting distribution in $B_{\rm fuv}$ is similar to that found in the single-IMF case but has a slightly higher median and {\em slightly} smaller scatter (P$_{84.1}-$P$_{15.9}=0.31$, c.f. P$_{84.1}-$P$_{15.9}=0.36$ in the single-IMF case). This suggests (for a fixed IMF) that the recent star formation history is the primary driver affecting the scatter in the UV-SFR calibration.

\subsubsection{Default Implementation}\label{sec:d.default}

\begin{figure}
\centering
\includegraphics[width=20pc]{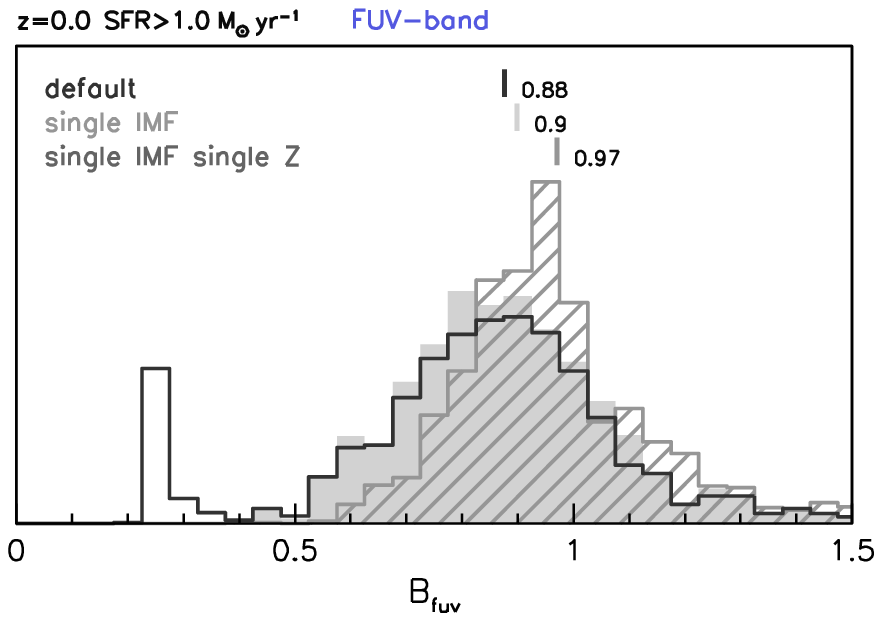}\\
\includegraphics[width=20pc]{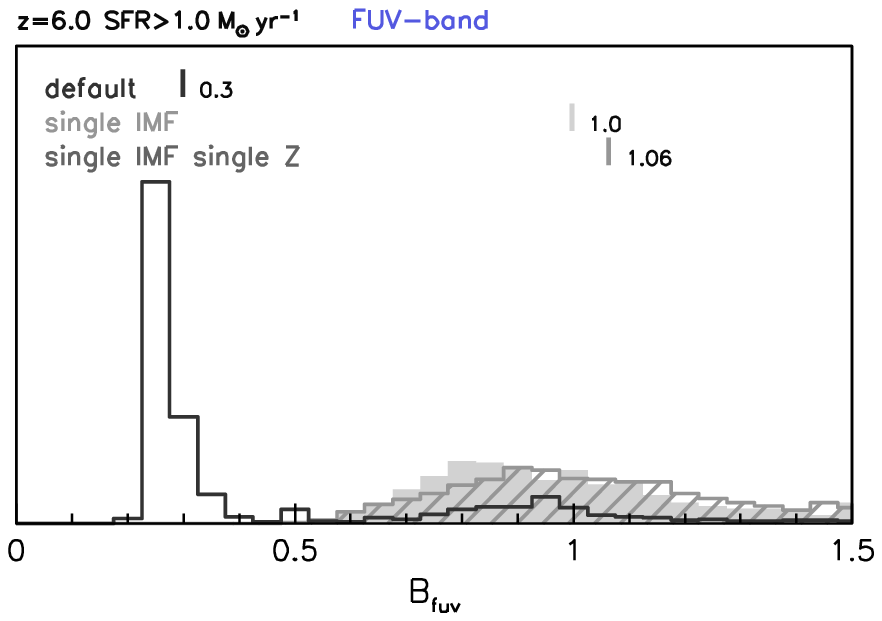}
\caption{The distribition of the UV-SFR calibrations ($B_{\rm fuv}$) for star forming galaxies (SFR$>1\,{\rm M_{\odot}\,yr^{-1}}$) at $z=0$ (top) and $z=6$ (bottom) assuming the {\em default} (line histogram), single IMF (shaded histogram), and single IMF single metallicity (hatched histogram) implementations of the B05 {\sc galform} model. The $y$-axis is arbitrary and on a linear scale. The bars show the median value of $B_{\rm fuv}$, which is written at the side. The histograms in each panel are normalised to contain the same number of galaxies.}
\label{fig:d_fuv}
\end{figure}

\begin{figure}
\centering
\includegraphics[width=20pc]{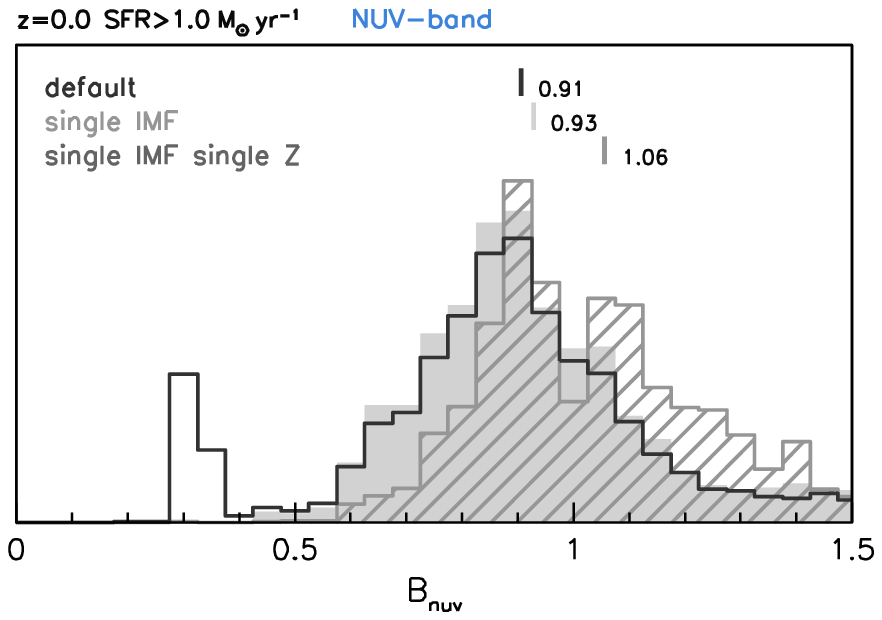}\\
\includegraphics[width=20pc]{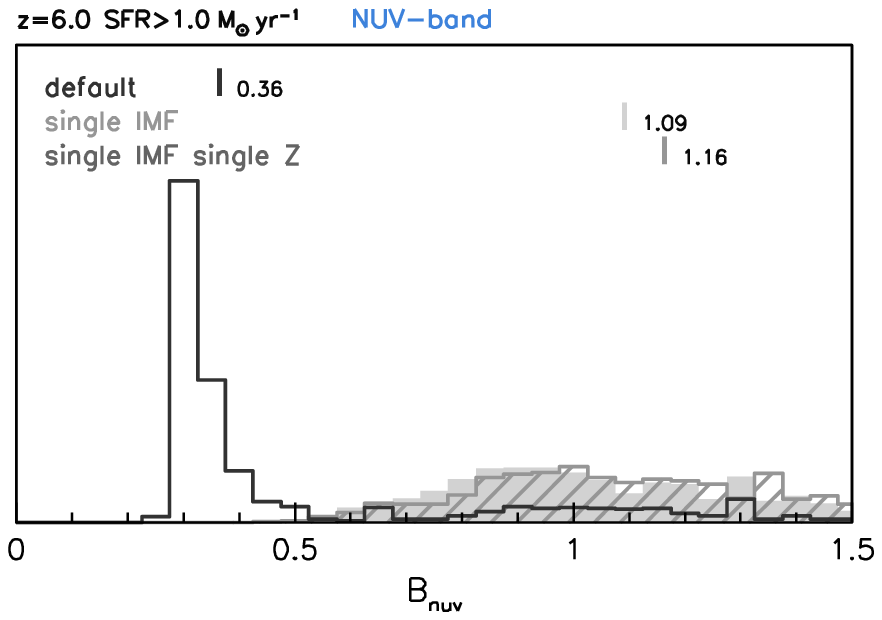}
\caption{The same as Fig. \ref{fig:d_fuv} but assuming the NUV-band filter.}
\label{fig:d_nuv}
\end{figure}

\begin{figure}
\centering
\includegraphics[width=20pc]{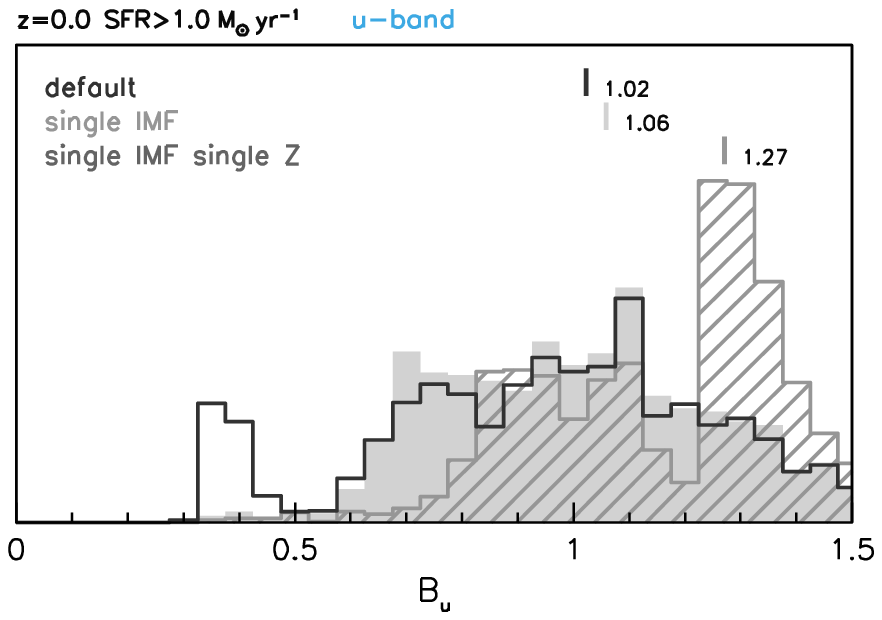}\\
\includegraphics[width=20pc]{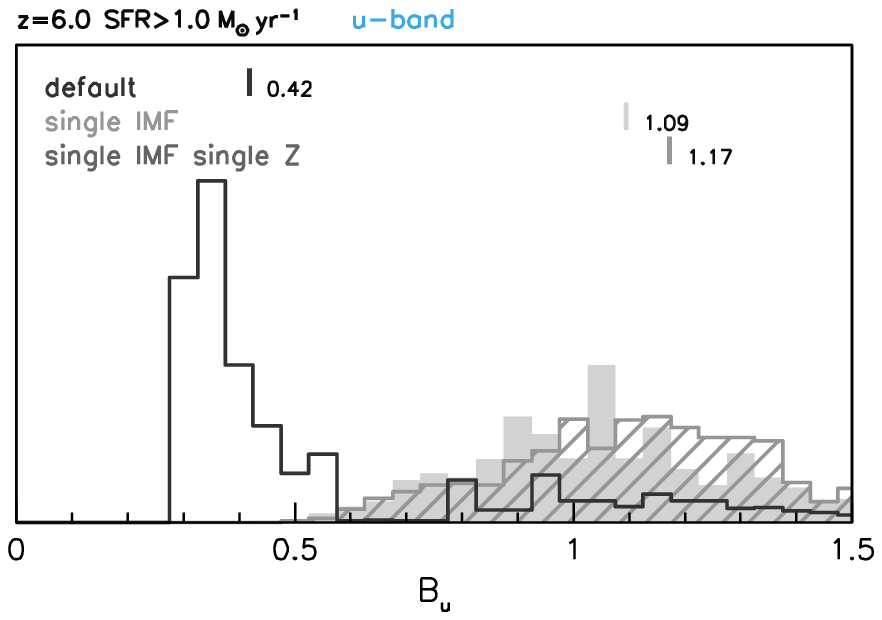}
\caption{The same as Fig. \ref{fig:d_fuv} but assuming the {\em u}-band filter.}
\label{fig:d_u}
\end{figure}

The distribution of $B_{\rm fuv}$ assuming the {\em default} implementation of the B05 model at low-redshift ($z=0$) has a similar median ($B_{\rm fuv}=0.88$, c.f. $B_{\rm fuv}=0.90$ for the single-IMF implementation) and profile to the single IMF case, with the exception of a small second peak around $B_{\rm fuv}=0.25$. This peak is due to merger driven star formation which, in the {\em default} implementation of the B05 model, occurs with a top-heavy IMF ($\xi\propto m^{-1}$). At high-redshift (lower-panel of Fig. \ref{fig:d_fuv}) much of the star formation in galaxies with SFR$>1\,M_{\odot}\,yr^{-1}$ is merger driven, and therefore occurs with a top-heavy IMF. This results in a strong peak at $B_{\rm fuv}=0.25$ and a lower amplitude distribution at $B_{\rm fuv}>0.5$. The resulting median of the distribution is $B_{\rm fuv}=0.30$, less than a third of that in the single-IMF implementation. This illustrates the difficulty in using the UV emission to infer the SFR in such a model, as $B_{\rm fuv}$ effectively becomes time dependent. 

\subsection{The predicted correlation with the intrinsic star formation rate}\label{sec:gp.sfr}

\begin{figure}
\centering
\includegraphics[width=20pc]{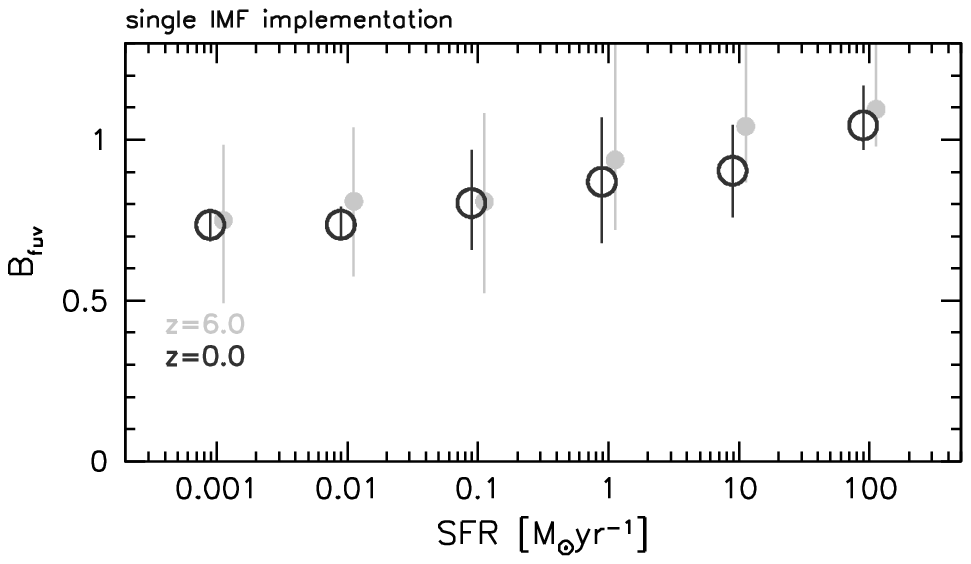}\\
\includegraphics[width=20pc]{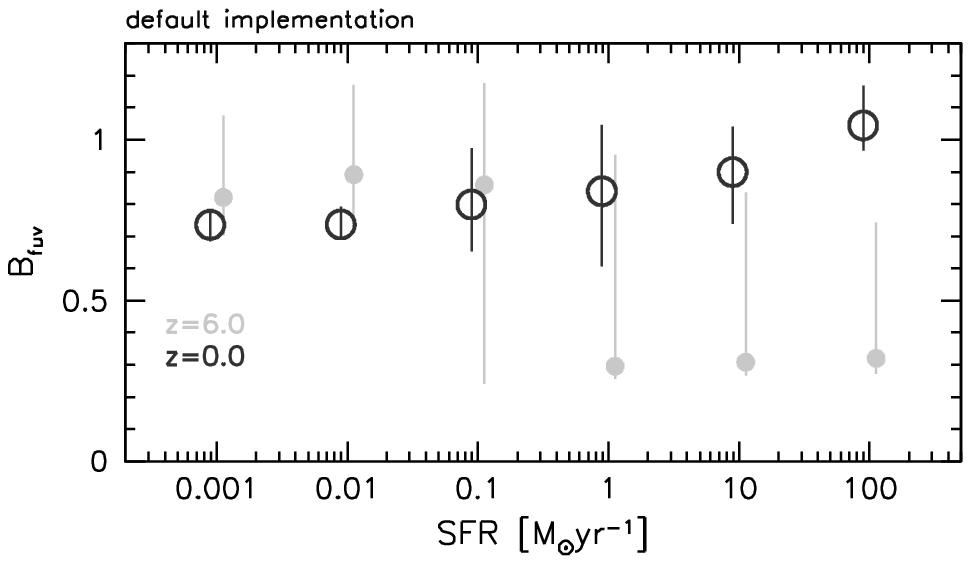}
\caption{The correlation of the median UV-SFR calibration $B_{\rm fuv}$ with star formation rate at $z=0$ (black) and $z=6$ (grey) assuming the single IMF model (top) and {\em default} implementation (bottom) of the B05 model. In each case the points denote the median value of $B_{\rm fuv}$ in each SFR bin while the lines denote the $68.2\%$ confidence interval (CI$_{\,68.2\%}$). The points are horizontally offset by $\pm 0.05\,{\rm dex}$ for clarity.}
\label{fig:sfr}
\end{figure}

In Fig. \ref{fig:sfr} the median UV-SFR calibration is shown for galaxies binned by star formation rate for the single IMF (top) and default implementations (bottom) of the B05 {\sc galform} model at $z=0$ and $z=6$.

In the single IMF implementation of the model there is, at both high and low-redshift, a correlation between SFR and the calibration. Galaxies with SFR$>1\,{\rm M_{\odot}\,yr^{-1}}$ have a median calibration $B_{\rm fuv}>0.9$ while those with SFR$<0.01\,{\rm M_{\odot}\,yr^{-1}}$ have $B_{\rm fuv}<0.8$. This is predominantly a result of the fact that galaxies with higher instantaneous star formation rates in the model typically have smaller UV-weighted ages, i.e. the contribution from lower-mass UV luminous stars is smaller. Crucially, this suggests that the naive application of a single calibration across all UV luminosities may result in the mis-estimation of the true SFR. Galaxies with star formation rates $\sim 100\,{\rm M_{\odot}\,yr^{-1}}$ appear to have FUV calibrations $\approx 20\%$ larger than the average for those with SFRs $\sim 1\,{\rm M_{\odot}\,yr^{-1}}$.

For the default implementation of the model (shown in the bottom panel of Fig. \ref{fig:sfr}) the behaviour is more complex. At low-redshift the correlation is similar to that seen in the single IMF implementation of the model; there is a weak correlation between SFR and $B_{\rm fuv}$. In contrast, at high-redshift the median value of calibration drops dramatically and the width of the distribution increases for SFR$>1\,{\rm M_{\odot}\,yr^{-1}}$ resulting in a large uncertainty. This is because at high-redshift, high star formation rates are dominated by merger driven star formation, which occurs with the flatter IMF which has a much lower calibration associated with it, as seen in \S\ref{sec:d.default}. 

\subsection{Correlation with redshift}\label{sec:gp.z}

\begin{figure}
\centering
\includegraphics[width=20pc]{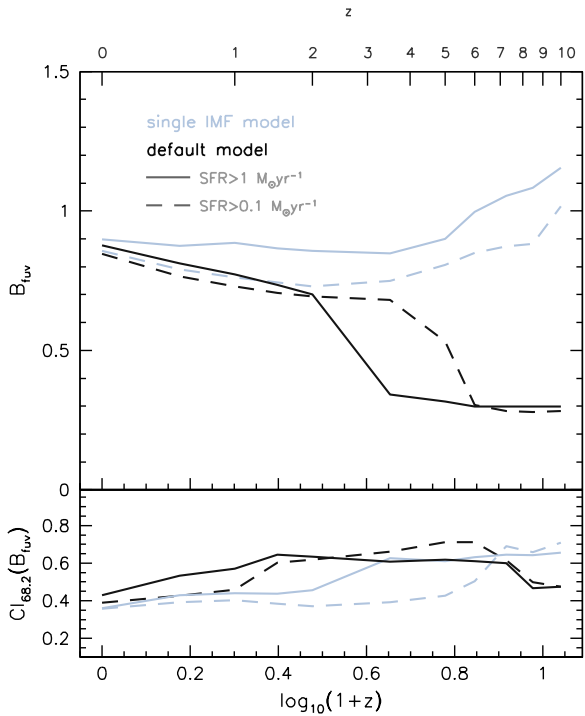}
\caption{The redshift evolution of the median UV-SFR calibration (top) and $68.2\%$ confidence interval width (bottom) for both the {\em default} (dark lines) and single IMF (light lines) implementations of the model assuming SFR$>1\,{\rm M_{\odot}\,yr^{-1}}$ (solid lines) and SFR$>0.1\,{\rm M_{\odot}\,yr^{-1}}$ (dashed lines).}
\label{fig:z}
\end{figure}

In Fig. \ref{fig:z} we show the evolution of both the median and $68.2\%$ confidence interval of the UV-SFR calibration $B_{\rm fuv}$ distribution using both the {\em default} and single IMF implementations of the {\sc galform} model for two star formation rate thresholds SFR$>0.1\,{\rm M_{\odot}\,yr^{-1}}$ and SFR$>1.0\,{\rm M_{\odot}\,yr^{-1}}$.

For the single IMF implementation and for galaxies with SFR$>1\,{\rm M_{\odot}\,yr^{-1}}$, reassuringly the median value of $B_{\rm fuv}$ remains roughly constant for redshifts $z=0\to 3$. However, at $z>4$ the calibration increases, climbing to $B_{\rm fuv}=1.1$ at $z=9$, as a result of the decreasing stellar ages in galaxies. This suggests that at very-high redshift, a larger calibration should be applied to the observed UV luminosities to correctly determine the star formation rate (and thus star formation rate density). The trend for galaxies with SFR$>0.1\,{\rm M_{\odot}\,yr^{-1}}$ is similar in form though at all redshifts the median is slightly smaller. In addition, the width of the distribution (as measured by the $68.2\%$ confidence interval) increases, reaching CI$_{68.2\%}\approx 0.6$ at very-high ($z>7$) redshift. 

Again, in the default implementation of the model for galaxies with SFR$>1\,{\rm M_{\odot}\,yr^{-1}}$ the behaviour is more complex. $B_{\rm fuv}$ declines slowly to $z=2$ before rapidly declining to $z=4$ and then flattening off at very-high redshift. This behaviour is driven by the evolving contribution of merger-driven star formation to the total star formation rate. For galaxies with SFR$>0.1\,{\rm M_{\odot}\,yr^{-1}}$ the trend is again similar though the transition to merger dominated SF takes place at slightly higher-redshift. 

\section{Conclusions}\label{sec:c}

The rest-frame ultraviolet (UV) luminosity of galaxies is widely used as a diagnostic of their instantaneous star formation rate. However, the calibration relating the UV luminosity to the star formation rate is sensitive to the recent star formation and metal enrichment history of the galaxy (as well as the choice of initial mass function). 

Using the {\sc galform} galaxy formation model to produce realistic star formation and metal enrichment histories we determine that the median calibration $B_{\rm fuv}$ (Eqn. \ref{eqn:calib}) is $\approx 0.9$ (assuming a Kennicutt 1983 IMF). This value is almost identical to that of a stellar population forming stars continuously for $100\,{\rm Myr}$ at solar metallicity (as is typically assumed in the literature to determine the calibration). However, there is a distribution of calibrations with a 68.2\% confidence interval of P$_{84.1}-$P$_{15.9}=0.36$. The width of this distribution implies, at $z=0$, for a single object, there is an uncertainty on the SFR as measured from the intrinsic FUV luminosity alone of $\approx 20\%$ (increasing to $\approx 23\%$ in the NUV and $\approx 34\%$ in the {\em u}-band) even in the absence of photometric noise, redshift uncertainty or dust. At higher-redshift this uncertainty increases becoming $\approx 30\%$ for the FUV-band at $z=6$ ($\approx 34\%$ and $\approx 32\%$ in the NUV and {\em u}-bands respectively).

We also investigate whether the recovered calibration $B_{\rm fuv}$ is correlated with star formation rate or redshift. Using a single-IMF implementation of the {\sc galform} model we find a weak positive correlation of $B_{\rm fuv}$ with SFR (irrespective of redshift) and a positive correlation with redshift (though only at $z>3$). If instead we use the {\em default} implementation of the Baugh et al. (2005) model, which adopts a top-heavy IMF in merger driven star formation, the situation is more complex. At high-redshift ($z>2$), where merger driven star formation dominates, the median calibration is $B_{\rm fuv}\approx 0.3$ reflecting the larger proportion of high-mass stars due to the top-heavy IMF.

Our results have implications for both theorists and observers. For simulators, it is apparent that the UV luminosity of a galaxy cannot be accurately determined from its instantaneous SFR alone, rather the spectral energy distribution should be computed by building a composite stellar population using the predicted star formation and metal enrichment history. Similarly, a single conversion from UV luminosity to SFR is only a rough approximation. A distribution of values should be adopted when interpreting observation data, whose median and width could well be functions of redshift and SFR.

\subsection*{Acknowledgements}

We would like to thank the anonymous referee for helpful suggestions which have greatly improved the paper. The calculations for this paper were performed on the ICC Cosmology Machine, which is part of the DiRAC Facility jointly funded by STFC, the Large Facilities Capital Fund of BIS, and Durham University. SMW acknowledges support from STFC. VGP acknowledges support from the UK Space Agency. VGP, CGL \& CMB acknowledge support from the Durham STFC rolling grant to the ICC.

\bsp

\end{document}